\documentclass[prd,aps,twocolumn,preprintnumbers,amsmath,amssymb,superscriptaddress]{revtex4}
\pdfoutput=1

\usepackage{amsmath}
\usepackage{graphicx}
\usepackage{dcolumn}
\usepackage{bm}
\usepackage{amssymb}
\usepackage{latexsym}
\usepackage{epstopdf}
\usepackage{threeparttable}
\usepackage{booktabs}
\usepackage{tabularx}
\usepackage{ulem}
\usepackage{float}
\usepackage{mathrsfs}
\usepackage{hhline}
\usepackage{multirow}
\usepackage{color}
\usepackage{subfigure}
\usepackage[colorlinks,linkcolor=magenta,anchorcolor=blue,citecolor=blue]{hyperref}

\makeatletter

\newcommand{\Rmnum}[1]{\expandafter\@slowromancap\romannumeral #1@}
\makeatother

\def\be{\begin{equation}}
	\def\ee{\end{equation}}
\def\bea{\begin{eqnarray}}
	\def\eea{\end{eqnarray}}

\begin{document}
\title{Cosmological constraints in covariant $f(Q)$ gravity with different connections }
	
\author{Jiaming Shi}
\email{shijiaming@ucas.ac.cn}

\affiliation{School of Fundamental Physics and Mathematical Sciences, Hangzhou Institute for Advanced Study,
University of Chinese Academy of Sciences, Hangzhou 310024, China}
\affiliation{School of Physical Sciences, University of Chinese Academy of Sciences, No.19A Yuquan Road, Beijing 100049, China}
\date{\today}
\begin{abstract}
Recently it has been shown that the cosmological dynamics of covariant $f(Q)$ gravity depend on different affine connections. In this paper, two specific $f(Q)$ models are investigated with SNe+CC+BAO+QSO observational data, and the spatial curvature of the universe is studied in covariant $f(Q)$ gravity. It is found that the parameters $\mathcal{X}_0$ and $\mathcal{X}'_0$ characterizing affine connections significantly affect the behavior of the effective equation of state $w_Q$ and may drive it across the phantom divide line. These results imply some inertial effects of the universe change the cosmic dynamics. However based on the Bayesian evidence, the zero inertial effect is more favored in the flat universe. Moreover, a closed universe is favored not only in the $\Lambda$CDM model but also in covariant $f(Q)$ gravity. The $f(Q)$ models have less support evidence than the $\Lambda$CDM model in the non-flat universe.
\end{abstract}

\maketitle

\section{introduction}
The covariant formulation of $f(Q)$ theory \cite{Hohmann:2019fvf,Zhao:2021zab,Hohmann:2021ast} extends the possibility of studying the dynamics of symmetric teleparallel gravity in the spherically symmetric coordinate system, especially in the case of cosmic spatial curvature \cite{Dimakis:2022rkd}. A self-consistent equation of motion does not exist in a general $f(Q)$ model if we apply the so-called coincident gauge with the usual partial derivatives for the definition of non-metricity instead of the covariant formulation of affine connections or the symmetric teleparallel covariant derivative. The coincident gauge in the spherically symmetric vacuum spacetime imposes a constraint equation on the expression of $f(Q)$ \cite{Zhao:2021zab}, i.e. $d^2f(Q)/dQ^2=0$, which lets $f(Q)$ theory goes back to Symmetric Teleparallel General Relativity (STGR) \cite{Nester:1998mp,BeltranJimenez:2017tkd}. It leads to the problem that the nonlinear function $f(Q)$ in the spherically symmetric
coordinate system is not compatible with the coincident gauge. A similar situation also happens in the modified Teleparallel Equivalent of General Relativity (TEGR) with vanishing non-metricity and zero curvature, such as $f(T)$ theory \cite{Krssak:2015oua,Krssak:2018ywd,Bahamonde:2021gfp}. Also, $f(Q)$ theory could separate inertial effects from gravitation unlike General Relativity (GR). It provides possibilities for studying inertial effects in gravitational systems.

Many works on $f(Q)$ theory in the applications of cosmology have been done, and these works use the coincident gauge which limits the investigation to the flat universe \cite{BeltranJimenez:2019tme,Lu:2019hra,Lazkoz:2019sjl,Mandal:2020buf,Ayuso:2020dcu,Frusciante:2021sio,
Anagnostopoulos:2021ydo,Atayde:2021pgb,Albuquerque:2022eac,Solanki:2022ccf,Anagnostopoulos:2022gej,Koussour:2022jyk,Ferreira:2022jcd,Arora:2022mlo,De:2022jvo,Lymperis:2022oyo,Narawade:2022cgb,
Khyllep:2022spx,Capozziello:2022wgl,Capozziello:2022tvv,Sokoliuk:2023ccw,Koussour:2023gip,Koussour:2023rly,Najera:2023wcw,Atayde:2023aoj,Ferreira:2023kgv,Narawade:2023nzv,Narawade:2023tnn}. $f(Q)$ gravity offers competitive alternatives to the concordance $\Lambda$CDM model. In addition to the cases of minimal coupling matter, the non-minimally coupled $f(Q)$ cosmology is proposed to explain the present cosmic accelerating expansion without introducing the cosmological constant \cite{Harko:2018gxr,Mandal:2021bpd}.
The $f(Q)$ models inducing the dynamical dark energy can naturally explain the phantom behavior ( the dark energy equation of state $w<-1$) in the late-time universe \cite{Lymperis:2022oyo,Narawade:2022cgb,Koussour:2023gip} and the possibility of crossing the phantom divide line $w=-1$ can be realized in some specific $f(Q)$ models \cite{Lu:2019hra,Arora:2022mlo,Solanki:2022ccf}.

As for the $f(Q)$ theory without the coincident gauge, it attracts more and more attention whether the cosmological investigation is in the flat universe  \cite{Subramaniam:2023okn,Shabani:2023nvm,Paliathanasis:2023nkb} or the curved one \cite{Dimakis:2022wkj,Dimakis:2022rkd,Heisenberg:2022mbo,Shabani:2023xfn,Subramaniam:2023old}. The new gauge choices could affect the dynamical behavior in cosmology since the Friedmann equations are modified \cite{Dimakis:2022rkd}. However, the evolution of the flat universe always experiences an unstable radiation-dominated era, then an unstable matter-dominated era, and finally a stable de Sitter stage in the $f(Q)$ theory with the nontrivial affine connection \cite{Shabani:2023nvm}. Hence the $f(Q)$ theory contributing the candidates for the dark energy can alleviate the cosmological constant problem. Furthermore, a suitable affine connection may lead to an early-time acceleration phase following the inflationary universe \cite{Paliathanasis:2023nkb}. Considering the case of non-vanishing spatial curvature, a curvature-dominated stage could happen in the early-time universe and the curvature density may exhibit a peak at intermediate times in the open universe \cite{Shabani:2023xfn}.

The goal of this paper is to investigate the existence of the nontrivial connections in the covariant $f(Q)$ theory by cosmological observations. Such a nontrivial connection can prove that the evolution of the universe is not only controlled by gravitation but also some inertial effects induced by this connection.

The paper is organized as follows. In Sec. \ref{Covariant}, I briefly introduce the covariant $f(Q)$ theory and list its modified Friedmann equations.   In Sec. \ref{Data}, I show the cosmological constraints on  the covariant $f(Q)$ theory with four different connections. This section separates into three parts as follows. Sec. \ref{dataA} introduces the data sets used in this paper. Sec. \ref{dataB} describes tow specific $f(Q)$ models and the MCMC results of the concordance $\mathrm{\Lambda}$CDM modeL and these two $f(Q)$ models . The model comparison is discussed in Sec. \ref{dataC}. Finally I draw the conclusions and give some discussion in Sec. \ref{Conclusions}.

\section{Covariant formulation of $f(Q)$ theory}
\label{Covariant}
The action of $f(Q)$ theory is given by \cite{Hohmann:2019fvf,Zhao:2021zab,Hohmann:2021ast}
\begin{align}\label{action}
S=\int d^{4}x\sqrt{-g}\left[\frac{1}{2}f(Q)+\mathcal{L}_{M}\right]\,,
\end{align}
where $\mathcal{L}_{M}$ is the matter Lagrangian density, $g$ is the determinant of the metric $g=\det(g_{\mu\nu})$, and $f(Q)$ is an arbitrary
function of the non-metricity scalar $Q$. The units $8\pi G=c=1$ are used in this paper. In metric-affine gravitational theories,  the non-metricity tensor $Q_{\alpha\mu\nu}$ defined by the basic dynamical objects $g_{\mu\nu}$
and  the connection $\Gamma_{~\mu\nu}^{\alpha}$,
\begin{align}\label{nm}
Q_{\alpha\mu\nu}=\nabla_{\alpha}g_{\mu\nu}=\partial_{\alpha}g_{\mu\nu}-\Gamma_{~\alpha\mu}^{\lambda}g_{\lambda\nu}-\Gamma_{~\alpha\nu}^{\lambda}g_{\lambda\mu}\,,
\end{align}
and  the curvature $R_{~\beta\mu\nu}^{\alpha}$ is given by
\begin{align}\label{curvature}
R_{~\beta\mu\nu}^{\alpha}=\frac{\partial\Gamma_{~\beta\nu}^{\alpha}}{\partial x^{\mu}}-\frac{\partial\Gamma_{~\beta\mu}^{\alpha}}{\partial x^{\nu}}+\Gamma_{~\sigma\mu}^{\alpha}\Gamma_{~\beta\nu}^{\sigma}-\Gamma_{~\sigma\nu}^{\alpha}\Gamma_{~\beta\mu}^{\sigma}\,,
\end{align}
here such theories have the flatness condition $R_{\beta\mu\nu}^{\alpha}=0$ and the torsionless condition $T_{~\mu\nu}^{\alpha}=\Gamma_{~\nu\mu}^{\alpha}-\Gamma_{~\mu\nu}^{\alpha}=0$. The vanishing curvature constraint forces the
connection to be purely inertial \cite{BeltranJimenez:2017tkd}. The theories with a flat and torsion-free geometry referring to the non-vanishing non-metricity tensor $Q_{\alpha\mu\nu}$ are called symmetric teleparallel theories since $Q_{\alpha\mu\nu}$ is symmetric in the last two indices \cite{Heisenberg:2018vsk}.  The non-metricity scalar is given by \cite{BeltranJimenez:2017tkd}
\begin{align}\label{ns}
Q=Q_{\alpha\mu\nu}P^{\alpha\mu\nu}\,,
\end{align}
where the non-metricity conjugate is
\begin{align}\label{P}
P_{~\mu\nu}^{\alpha}=-\frac{1}{2}L_{~\mu\nu}^{\alpha}+\frac{1}{4}\left(Q^{\alpha}-\overline{Q}^{\alpha}\right)g_{\mu\nu}-\frac{1}{4}\delta_{(\mu}^{\alpha}Q_{\nu)}\,,
\end{align}
with the  disformation
\begin{align}\label{df}
L_{~\mu\nu}^{\alpha}=\frac{1}{2}Q_{~\mu\nu}^{\alpha}-Q_{(\mu~\nu)}^{~~\alpha}\,,
\end{align}
and the traces $Q_{\alpha}=Q_{\alpha\mu\nu}g^{\mu\nu}$, $\overline{Q}_{\alpha}=Q_{\mu\nu\alpha}g^{\mu\nu}$. Here the
parentheses in the indices denote the symmetrization of tensor, i.e. $A_{(\mu\nu)}=\frac{1}{2}\left(A_{\mu\nu}+A_{\nu\mu}\right)$.
Two kinds of equations of motion can be obtained in the symmetric teleparallel theory since the action \eqref{action} is constructed by the metric and the connection independent from the metric. The equations of motion for the metric are
\begin{align}\label{eom1}
f_{Q}G_{\mu\nu}+\frac{1}{2}g_{\mu\nu}(f_{Q}Q-f)+2f_{QQ}\nabla_{\alpha}QP_{~~\mu\nu}^{\alpha}=T_{\mu\nu}\,,
\end{align}
where $G_{\mu\nu}$ is the Einstein tensor and  $T_{\mu\nu}$ is the energy-momentum tensor of matter. Here I denote $f_{Q}=\partial f/\partial Q,f_{QQ}=\partial^{2}f/\partial Q^{2}$.
Variation with respect to the connection, the equations of motion are derived as
\begin{align}\label{eom2}
\nabla_{\mu}\nabla_{\nu}\left(\sqrt{-g}f_{Q}P_{~~~\sigma}^{\mu\nu}\right)=0\,.
\end{align}
Eq.\eqref{eom2} corresponds to the conservation law $\mathring{\nabla}_{\mu}T_{~~\nu}^{\mu}=0$  for the matter energy-momentum tensor \cite{Harko:2018gxr,BeltranJimenez:2018vdo,De:2022jvo}. The symbol $\mathring{\nabla}$ is used to denote the covariant derivative with respect to the Christoffel symbols $\mathring{\Gamma}_{~\mu\nu}^{\alpha}$ defined as
\begin{align}\label{cf}
\mathring{\Gamma}_{~\mu\nu}^{\alpha}=\frac{1}{2}g^{\alpha\lambda}\left(\partial_{\mu}g_{\nu\lambda}+\partial_{\nu}g_{\mu\lambda}-\partial_{\lambda}g_{\mu\nu}\right)\,.
\end{align}
 The general affine connection without torsion can be related to the Christoffel connection by
\begin{align}\label{GGL}
\Gamma_{~\mu\nu}^{\alpha}=\mathring{\Gamma}_{~\mu\nu}^{\alpha}+L_{~\mu\nu}^{\alpha}\,.
\end{align}
In Eq. \eqref{GGL}, the Christoffel connection $\mathring{\Gamma}_{~\mu\nu}^{\alpha}$ mixing the inertia and the gravitation \cite{Aldrovandi:2013wha,Capozziello:2022zzh} which reflects the Einstein Equivalence Principle can be decomposed into the inertial effects  $\Gamma_{~\mu\nu}^{\alpha}$ as the non-covariant part and the gravitational part $L_{~\mu\nu}^{\alpha}$ as the covariant force \cite{BeltranJimenez:2019esp}. It could deduce a gravitational analog of the Lorentz force equation in geodesic equation \cite{deAndrade:1997gka}. Indeed the affine connection just as pure inertial effects can be parameterized by some arbitrary functions $\xi^{\lambda}(x)$ so that $\Gamma_{~\mu\nu}^{\alpha}=\left(\partial x^{a}/\partial\xi^{\lambda}\right)\partial_{\mu}\partial_{\nu}\xi^{\lambda}$ \cite{BeltranJimenez:2017tkd}, it differs from the
trivial connection by a general linear gauge transformation. In the language of gauge theory, what is shown by Eq. \eqref{GGL} is similar to the spin gauge field $\mathcal{A}_{\mu}^{ab}$ being decomposed into spin graviguage field $\Omega_{\mu}^{ab}$ and spin covariant gauge field $\mathrm{A}_{\mu}^{ab}$, as expressed by  $\mathcal{A}_{\mu}^{ab}=\Omega_{\mu}^{ab}+\mathrm{A}_{\mu}^{ab}$ \cite{Wu:2022mzr}. Here the spin graviguage field $\Omega_{\mu}^{ab}$ like the Christoffel connection induces the geometrical structure of the action in GR.

Let's move on to the covariant formulation of the symmetric teleparallel theory. Since coincident gauge $\nabla_{\alpha}g_{\mu\nu}=\partial_{\alpha}g_{\mu\nu}$, i.e. $\Gamma_{~\mu\nu}^{\alpha}=0$ is incompatible with the spherically symmetric spacetime, we need to generalize the connection to represent the inertial effects in the gravitational system we are interested in. That means the connection $\Gamma_{~\mu\nu}^{\alpha}$ remains the same whether the gravity is canceled or not, as expressed by \cite{Lin:2021uqa}
\begin{align}\label{gg}
\Gamma_{~\mu\nu}^{\alpha}=\left.\Gamma_{~\mu\nu}^{\alpha}\right|_{G=0}=\left.\mathring{\Gamma}_{~\mu\nu}^{\alpha}\right|_{G=0}\,.
\end{align}
Thus it is easy to find that the coincident gauge $\Gamma_{~\mu\nu}^{\alpha}=0$ is compatible with the Cartesian coordinate system where $\left.\mathring{\Gamma}_{~\mu\nu}^{\alpha}\right|_{G=0}=0$, which means no inertial effects exiting in the gravitational system. However, Eq. \eqref{gg} can not work in cosmology since we cannot cancel gravity directly to get a Minkowski vacuum as the universe is evolving. In order to study the cosmological dynamics of symmetric teleparallel gravity theories in  Friedmann-Lema\^{\i}tre-Robertson-Walker (FLRW) metric,
\begin{align}\label{FRLW}
ds^{2}=-dt^{2}+a(t)^{2}\left[\frac{dr^{2}}{1-kr^{2}}+r^{2}\left(d\theta^{2}+\sin^{2}\theta d\phi^{2}\right)\right]\,,
\end{align}
an affine connection demands its Lie derivative along $X^\mu$ vanishes, i.e. $\left(\mathscr{L}_{X}\Gamma\right)_{\mu\nu}^{\alpha}=0$, and the vector $X^\mu$ is the Killing vector in the spacetime $\left(\mathscr{L}_{X}g\right)_{\mu\nu}=0$. Applying the definitions of their Lie derivatives
\begin{align}\label{lg}
\left(\mathscr{L}_{X}g\right)_{\mu\nu}=X^{\alpha}\partial_{\alpha}g_{\mu\nu}+\partial_{\nu}X^{\alpha}g_{\alpha\nu}+\partial_{\nu}X^{\alpha}g_{\mu\alpha}\,,
\end{align}
\begin{align}\label{LC}
\left(\mathscr{L}_{X}\Gamma\right)_{\mu\nu}^{\alpha}=&X^{\lambda}\partial_{\lambda}\Gamma_{\mu\nu}^{\alpha}
-\partial_{\lambda}X^{\alpha}\Gamma_{\mu\nu}^{\lambda}+\partial_{\mu}X^{\lambda}\Gamma_{\lambda\nu}^{\alpha}\\ \nonumber
&+\partial_{\nu}X^{\lambda}\Gamma_{\mu\lambda}^{\alpha}+\partial_{\mu}\partial_{\nu}X^{\alpha}\,,
\end{align}
we can get six spatial Killing symmetries $X^\mu$ and the following nonzero components of the torsionless connection
\cite{Hohmann:2020zre,Hohmann:2021ast,Heisenberg:2022mbo}
\begin{align}\label{GeneralC}
 \Gamma_{~tt}^{t}&=K_{1}\,,~~\Gamma_{~rr}^{t}=\frac{K_{2}}{\chi^{2}}\,,~~\Gamma_{~\theta\theta}^{t}=K_{2}r^{2}\,, \nonumber \\
\Gamma_{~tr}^{r}&=\Gamma_{~rt}^{r}=\Gamma_{~t\theta}^{\theta}=\Gamma_{~\theta t}^{\theta}=\Gamma_{~t\phi}^{\phi}=\Gamma_{~\phi t}^{\phi}=K_{3} \,,\nonumber\\
\Gamma_{~r\theta}^{\theta}&=\Gamma_{~\theta r}^{\theta}=\Gamma_{~r\phi}^{\phi}=\Gamma_{~\phi r}^{\phi}=\frac{1}{r}\,,~~\Gamma_{~\theta\theta}^{r}=-r\chi^{2}\,, \nonumber\\
~~\Gamma_{~\phi\phi}^{r}&=-r\chi^{2}\sin^{2}\theta\,,~~\Gamma_{~\phi\phi}^{t}=K_{2}r^{2}\sin^{2}\theta\,,\nonumber\\
\Gamma_{~\phi\theta}^{\phi}&=\Gamma_{~\theta\phi}^{\phi}=\cot\theta\,,~~\Gamma_{~\phi\phi}^{\theta}=-\sin\theta\cos\theta\,,
\end{align}
where  $K_1(t),K_2(t),K_3(t)$ are functions of time and $\chi^{2}=1-kr^{2}$. Using Eqs. \eqref{nm} and \eqref{ns}, the non-metricity scalar can be obtained, which reads
\begin{align}\label{QFRLW}
Q=&-6H^{2}+3K_{3}(K_{1}-K_{3})-\frac{3K_{2}(K_{1}+K_{3})}{a^{2}}\nonumber\\&+9K_{3}+\frac{3K_{2}H}{a^{2}}\,,
\end{align}
where $H=\dot{a}/a$ is the Hubble parameter, here the dot '.' denotes the derivative with respect to the cosmic time $t$. The non-vanishing components of  the disformation tensor are
\begin{align}\label{Labc}
&L_{~tt}^{t}=K_{1}\,, L_{~rr}^{t}=-\frac{a^{2}H-K_{2}}{\chi^{2}}\,,\nonumber\\ &L_{~\theta\theta}^{t}=-r^{2}\left(a^{2}H-K_{2}\right)\,,
L_{~\phi\phi}^{t}=-r^{2}\sin^{2}\theta\left(a^{2}H-K_{2}\right)\,,\nonumber\\
&L_{~tr}^{r}=L_{~rt}^{r}=L_{~t\theta}^{\theta}=L_{~\theta t}^{\theta}=L_{~t\phi}^{\phi}=L_{~\phi t}^{\phi}=K_{3}-H\,.
\end{align}
Finally, the vanishing curvature $R_{~\beta\mu\nu}^{\alpha}=0$ leads to the following constraint equations
\begin{align}\label{Keq}
K_{3}(K_{1}-K_{3})-\dot{K}_{3}=K_{2}(K_{1}-K_{3})+\dot{K}_{2}=k+K_{2}K_{3}=0\,.
\end{align}
From Eq.\eqref{Keq}, the solutions to $K_1(t),K_2(t),K_3(t)$ are classified to four cases:
\begin{enumerate}
  \item Connection $\Gamma_1$: $k=0,K_{1}=\gamma,K_{2}=0,K_{3}=0,$
  \item Connection $\Gamma_2$: $k=0,K_{1}=\frac{\dot{\gamma}}{\gamma}+\gamma,K_{2}=0,K_{3}=\gamma,$
  \item Connection $\Gamma_3$: $k=0,K_{1}=-\frac{\dot{\gamma}}{\gamma},K_{2}=\gamma,K_{3}=0,$
  \item Connection $\Gamma_4$: $K_{1}=-\frac{k+\dot{\gamma}}{\gamma},K_{2}=\gamma,K_{3}=-\frac{k}{\gamma}.$
\end{enumerate}
Here $\gamma(t)$ is a function of time. Clearly Connection $\Gamma_4$ goes back to  Connection $\Gamma_3$ if $k=0$, hence we can study the spatial curvature of the universe in one uniform expression as Connection $\Gamma_4$ whatever $k$ is.

In the FRLW metric, four categories of the modified Friedmann equations from Eq. \eqref{eom1} can be obtained corresponding to four connections \cite{Dimakis:2022rkd}.
For Connection $\Gamma_1$, the non-metricity scalar $Q$  and the Friedmann equations are
\begin{align}\label{Q1}
 &Q=-6H^{2}\,,\\
 \label{F1}
 &3H^{2}f_{Q}+\frac{1}{2}\left(f-Qf_{Q}\right)=\rho\,,\\
 \label{F12}
& -2\frac{d\left(f_{Q}H\right)}{dt}-3H^{2}f_{Q}-\frac{1}{2}\left(f-Qf_{Q}\right)=p\,,
\end{align}
where $\rho$ is the energy density of matter in the universe including baryonic matter, cold dark matter and radiation. $p$ is the pressure of the fluid. Here $f_Q=df/dQ$ and $f_{QQ}=d^2f/dQ^2$.
For Connection $\Gamma_2$, they are
\begin{align}\label{Q2}
&Q=-6H^{2}+9\gamma H+3\dot{\gamma}\,,\\
&3H^{2}f_{Q}+\frac{1}{2}\left(f-Qf_{Q}\right)+\frac{3\gamma}{2}\dot{Q}f_{QQ}=\rho\,,\\
&-2\frac{d\left(f_{Q}H\right)}{dt}-3H^{2}f_{Q}-\frac{1}{2}\left(f-Qf_{Q}\right)+\frac{3\gamma}{2}\dot{Q}f_{QQ}=p\,.
\end{align}
For Connection $\Gamma_3$, they are
\begin{align}\label{Q2}
&Q=-6H^{2}+\frac{3\gamma H}{a^{2}}+\frac{3\dot{\gamma}}{a^{2}}\,,\\
&3H^{2}f_{Q}+\frac{1}{2}\left(f-Qf_{Q}\right)-\frac{3\gamma}{2a^{2}}\dot{Q}f_{QQ}=\rho\,,\\
&-2\frac{d\left(f_{Q}H\right)}{dt}-3H^{2}f_{Q}-\frac{1}{2}\left(f-Qf_{Q}\right)+\frac{\gamma}{2a^{2}}\dot{Q}f_{QQ}=p\,.
\end{align}
For Connection $\Gamma_4$, they are
\begin{widetext}
\begin{align}\label{Q2}
&Q=-6H^{2}+\frac{3\gamma H}{a^{2}}+\frac{3\dot{\gamma}}{a^{2}}+k\left[\frac{6}{a^{2}}+\frac{3}{\gamma}\left(\frac{\dot{\gamma}}{\gamma}-3H\right)\right]\,,\\
&3H^{2}f_{Q}+\frac{1}{2}\left(f-Qf_{Q}\right)-\frac{3\gamma}{2a^{2}}\dot{Q}f_{QQ}+3k\left(\frac{f_{Q}}{a^{2}}-\frac{\dot{Q}f_{QQ}}{2\gamma}\right)=\rho\,,\\
&-2\frac{d\left(f_{Q}H\right)}{dt}-3H^{2}f_{Q}-\frac{1}{2}\left(f-Qf_{Q}\right)+
\frac{\gamma}{2a^{2}}\dot{Q}f_{QQ}-k\left(\frac{f_{Q}}{a^{2}}+\frac{3\dot{Q}f_{QQ}}{2\gamma}\right)=p\,.
\end{align}
\end{widetext}
The function $\gamma$ does not affect Eqs. \eqref{Q1}-\eqref{F12}, which is the same as the result in the flat FLRW metric of Cartesian coordinates $ds^2=-dt^2+a(t)^2d\mathbf{x}^2$ with the coincident gauge $\Gamma_{~\mu\nu}^{\alpha}=0$. In the study of the cosmological aspect, many works usually choose the case of Connection $\Gamma_1$. Therefore to investigate $f(Q)$ theory, it is very interesting to study the cosmological effects of the distinct connections. It helps us get to know what inertial effects exist in our universe.
\section{Data and methodology }
\label{Data}
In this section, I apply the cosmological probes to constrain $f(Q)$ theory, including SNe, CC, BAO and QSOs.
Now some useful variables to relate the theory with the observations need to be introduced. The transverse comoving distance $D_{M}(z)$ from the source to us is
\begin{align}\label{DM}
D_{M}(z)=\frac{c}{H_{0}\sqrt{\left|\Omega_{k}\right|}}\mathrm{sinn}\left(H_{0}\sqrt{\left|\Omega_{k}\right|}\int_{0}^{z}\frac{dz'}{H(z')}\right)\,,
\end{align}
and $\mathrm{sinn}(x)$ is defined as
\begin{align}\label{sinn}
\mathrm{sinn}(x)=\begin{cases}
\sin(x), & \Omega_{k}<0\,,\\
x, & \Omega_{k}=0\,,\\
\sinh(x), & \Omega_{k}>0\,,
\end{cases}
\end{align}
where $\Omega_{k}=-k/a_{0}^{2}H_{0}^{2}$ with the  scale factor $a_0=1$ for the current universe, $c$ is the speed of light and $H_0$ is the Hubble parameter today.
The luminosity distance can be obtained by
\begin{align}\label{DL0}
D_{L}(z)=(1+z)D_{M}(z)\,.
\end{align}
In fact, considering the peculiar velocity of the observer, the luminosity distance of SNe is defined as \cite{Davis:2010jq}
\begin{align}\label{DL}
D_{L}^{\mathrm{SNe}}=(1+z_{\mathrm{hel}})D_{M}(z_{\mathrm{cmb}})\,,
\end{align}
where $z_{\mathrm{cmb}}$ is the CMB restframe redshifts of SNe and $z_{\mathrm{hel}}$ is the heliocentric redshifts. In the following I introduce the data sets and describe the main results on two different $f(Q)$ models.
\subsection{Data}
\label{dataA}
\subsubsection{SNe}
The sample of Type Ia supernovae (SNe Ia) is called Pantheon Sample. This sample contains 1048 sources in the redshift range $0.01<z<2.26$ \cite{Pan-STARRS1:2017jku}.
This sample covers the data records from the Pan-STARRS1 (PS1) Medium Deep Survey, Sloan Digital Sky Survey (SDSS), Supernova Legacy Survey (SNLS), and Hubble Space Telescope (HST) survey. The standard description provides numerical values of the distance modulus, which can be directly employed to derive the luminosity distance $D_{L}$ (in Mpc) according to:
\begin{align}\label{SNe1}
\mu(z)=5\log_{10}\left[\frac{D_{L}^{\mathrm{SNe}}}{\mathrm{Mpc}}\right]+25\,.
\end{align}
The apparent magnitude $m$ of a supernova at redshift $z$ is given by
\begin{align}\label{SNe2}
m=\mu+M_B\,,
\end{align}
 where $M_B$ is the absolute magnitude. The chi-square ($\chi^2$) of SNe Ia is written as follows

\begin{align}\label{SNe3}
\chi_{\mathrm{SNe}}^{2}=\sum_{i=1}^{1048}\frac{\left(m_{th}(z_{i})-m_{obs}(z_{i})\right)^{2}}{\sigma_{m}^{2}(z_{i})}\,.
\end{align}
Here $\sigma_{m}$ is the observed error of the apparent magnitude $m$, while $m_{th}$ and $m_{obs}$ are the theoretical value and the observational value, respectively.
\subsubsection{CC}
The Hubble parameter $H(z)$ can be estimated at certain redshifts $z$ by
\begin{align}\label{cc}
H(z)=\frac{\dot{a}}{a}=-\frac{1}{1+z}\frac{dz}{dt}\simeq-\frac{1}{1+z}\frac{\Delta z}{\Delta t}\,.
\end{align}
Determining $\Delta z$ via a spectroscopic survey and differential ages $\Delta t$ (DA method) of passively evolving galaxies \cite{Jimenez:2001gg,Simon:2004tf}, it is possible to obtain the value of $H(z)$.
Compilations of such observations can be regarded as cosmic chronometers (CC), and I use a sample of 31 objects covering the redshift range $0<z<1.97$  \cite{Marra:2017pst}. For these measurement one can construct a $\chi_{\mathrm{CC}}^{2}$ estimator as follows:
\begin{align}\label{CC1}
\chi_{\mathrm{CC}}^{2}=\sum_{i=1}^{31}\frac{\left(H_{th}(z_{i})-H_{obs}(z_{i})\right)^{2}}{\sigma_{H}^{2}(z_{i})}\,.
\end{align}
Here, $H_{obs}$  and $H_{th}$ represent the observational value with its error $\sigma_{H}$ and the theoretical value of the Hubble parameter.
\subsubsection{BAO}
The theoretical BAO angular scale $\theta(z)$ can be written in terms of the angular diameter distance $D_{A}=aD_{M}$
\begin{align}\label{BAO1}
\theta(z)=\frac{180^{\circ}}{\pi}\frac{r_{s}a}{D_{A}}=\frac{180^{\circ}}{\pi}\frac{r_{s}}{D_{M}}\,,
\end{align}
and $r_{s}$ (in Mpc) is the sound horizon of the primordial photon-baryon fluid.
Here 14 data points from BAO data sets including SDSS-DR7 \cite{Alcaniz:2016ryy}, SDSS-DR10 \cite{Carvalho:2015ica}, SDSS-DR11 \cite{Carvalho:2017tuu}, SDSS-DR12Q \cite{deCarvalho:2017xye} are used.
The $\chi_{BAO}^{2}$ estimator for BAO is defined in the following manner
\begin{align}\label{BAO2}
\chi_{\mathrm{BAO}}^{2}=\sum_{i=1}^{14}\frac{\left(\theta_{th}(z_{i})-\theta_{obs}(z_{i})\right)^{2}}{\sigma_{\theta}^{2}(z_{i})}\,.
\end{align}
Here, $\theta_{obs}$  and $\theta_{th}$ represent the observational value with its error $\sigma_{\theta}$ and the theoretical value.
\subsubsection{QSOs}

Quasars or quasi-stellar objects (QSOs) are astrophysical objects of very high luminosity regarded as active galactic nuclei (AGN).
The quasar sample RL19 \cite{Risaliti:2018reu} consists of 1598 objects in the redshift
range $0.04<z<5.1$ with high-quality UV and X-ray flux measurements. Quasars as high-redshift standard candles were investigated in
Refs. \cite{Risaliti:2018reu,Lusso:2017hgz,Lusso:2019akb,Salvestrini:2019thn,Lusso:2020pdb,Sacchi:2022ofz} and it is suggested that Hubble diagrams of quasars have the $\sim4\sigma$ deviation from the $\Lambda$CDM model, which is not due to unknown
systematic effects. However, this deviation can be reduced via different data modeling methods and Monte Carlo Markov Chain (MCMC) implements \cite{Hu:2022udt}. It is considered that the X-rays are produced by a plasma of hot relativistic electrons through inverse Compton scattering processes on the seed UV photons \cite{Risaliti:2015zla,Salvestrini:2019thn}.
There exists a nonlinear relation between the luminosities in the X-rays ($L_{X}$) and UV band ($L_{UV}$ )
\begin{align}\label{QSO1}
\log(L_{X})=\alpha\log(L_{UV})+\beta\,.
\end{align}
This relation shows the stable physical property of quasars since the slope $\alpha$ is almost a constant at all redshifts \cite{Lusso:2020pdb,Risaliti:2023uiy}.
The flux $F$ and luminosity $L$ satisfy
\begin{align}\label{LF}
L=4\pi D_{L}^{2}F\,,
\end{align}
where $D_{L}$ is luminosity distance. Thus the following equation can be obtained
\begin{align}\label{QSO2}
\log F_{X}=\alpha\log F_{UV}+(2\alpha-2)\log_{10}D_{L}+\hat{\beta}\,,
\end{align}
with $\hat{\beta}=\beta+\left(\alpha-1\right)\log4\pi$. Here, $F_{X}$ and $F_{UV}$ represent the X-ray and UV flux, respectively.
The intrinsic dispersion $\delta$ of the $L_X-L_{UV}$ relation is considered to reduce the Eddington bias which has the effect of flattening the $L_X-L_{UV}$ relation \cite{Lusso:2020pdb}
\begin{align}\label{QSO3}
s_{i}^{2}=\sigma_{\log(F_{X})}^{2}+\alpha^{2}\sigma_{\log(F_{UV})}^{2}+\delta^{2}\,.
\end{align}
The likelihood
function or modified chi-square function for $F_X$ including a penalty term for the intrinsic dispersion $\delta$ is defined as
\begin{align}\label{QSO4}
\chi_{\mathrm{QSO}}^{2}=\sum_{i=1}^{1598}\left[\frac{\left[\log(F_{X,i}^{th})-\log(F_{X,i}^{obs})\right]^{2}}{s_{i}^{2}}+\ln(2\pi s_{i}^{2})\right]\,.
\end{align}

\subsection{Models and results}
\label{dataB}
In this section, results of constraints from observational data SNe+CC+BAO+QSO on the covariant $f(Q)$ theory are displayed and discussed. In order to use the data sets to constrain $f(Q)$ theory with different connections, two $f(Q)$ models are given by the following expressions:
\begin{align}\label{fQ1}
f(Q)=Qe^{\lambda Q_{0}/Q}\,,
\end{align}
which is dubbed as Exp-$f(Q)$ \cite{Anagnostopoulos:2021ydo,Lymperis:2022oyo,Ferreira:2023kgv} and
\begin{align}\label{fQ2}
f(Q)=Q+\lambda Q_{0}^{2}/Q\,,
\end{align}
which is dubbed as Inv-$f(Q)$ model \cite{BeltranJimenez:2019tme,Ayuso:2020dcu,Narawade:2023nzv} since additional term is the inverse of $Q$. Here $Q_0=Q(z=0)$ for the current universe.  Two models both go back to STGR or equivalently recover GR but not $\Lambda$CDM when $\lambda=0$, thus the additional modification to STGR may alleviate the cosmological constant problem since the case of $\lambda\neq0$ opens the door to a de Sitter phase in the future universe \cite{BeltranJimenez:2019tme,Ferreira:2023kgv}. The modified Friedmann equations can be uniformly written as
\begin{align}\label{modF21}
&3H^{2}=\rho_{N}+\rho_{Q}\,, \\
\label{modF22}
&-2\dot{H}=\rho_{N}+p_{N}+(1+w_{Q})\rho_{Q}\,,
\end{align}
where $\rho_N$ represents the energy density of contents in the universe $\rho_{N}=\rho_{m}+\rho_{r}+\rho_{k}$ with $\rho_{m}$ is the energy density of non-relativistic matter (baryonic matter, cold dark matter) and $\rho_{r}$ is that of radiation. Here $\rho_{k}=3H_{0}^{2}\Omega_{k}a^{-2}$  as the curvature energy density. $p_{N}=p_{r}+p_{k}$ denotes the pressure from radiation and curvature given by $p_{k}=-\rho_{k}/3$, $p_{r}=\rho_{r}/3$. $\rho_{Q}$ is the effective energy density of dark energy with the equation of state (EoS) $w_Q$, which is induced by the gravitational modifications in $f(Q)$ theory. The first modified Friedmann equation can be expressed by density parameters
\begin{align}\label{modF3}
3H^{2}=3H_{0}^{2}\left[\Omega_{m}(1+z)^3+\Omega_{r}(1+z)^4+\Omega_{k}(1+z)^2+\Omega_{Q}(z)\right]\,,
\end{align}
where the density parameters $\Omega_{i}(z)=\rho_{i}(z)/3H^{2}$ with the index $i=m,r,k,Q$ representing non-relativistic matter, radiation, curvature and dark energy, respectively. Note $\Omega_Q(z)$ is not a constant in Eq. \eqref{modF3} while $\Omega_{m},\Omega_{r},\Omega_{k}$ are constant standing for energy density fractions in the current universe. Also, $H_{0}$ represents the Hubble constant today. From Eqs. \eqref{modF21}\eqref{modF22}, one can obtain $w_Q$ by knowing the evolution of $H(z)$ and the values of density parameters $\Omega_{i}$,
\begin{align}\label{wQ}
w_{Q}(z)=\frac{-2\dot{H}-3H^{2}-p_{N}(z)}{3H^{2}-\rho_{N}(z)}\,.
\end{align}
Now I set $\Omega_{r}=4.184\times10^{-5}/h^2 $ \cite{10.1093/oso/9780198570899.001.0001} with the definition $H_0=100h ~\mathrm{km/s/Mpc}$, while $\Omega_{m},\Omega_{k}$ are determined by the observational constraints in the $f(Q)$ cosmology. Thus $p_{N}(z),\rho_{N}(z)$ can be computed, and then the value of $w_Q$ at any redshift will be deduced. To investigate the effect of the non-vanishing $\gamma$ in the affine connections Eq.\eqref{GeneralC} to $f(Q)$ cosmology, I set initial conditions for $\gamma$ at $z=0$:
\begin{align}\label{gamma0}
&\gamma(z=0)= \mathcal{X}_0 ~\mathrm{km/s/Mpc}\,,\\
\label{dgammadx}
&\frac{d\gamma}{d\ln a}(z=0)= \mathcal{X}'_0 ~\mathrm{km/s/Mpc}\,.
\end{align}
In fact Eqs. \eqref{gamma0}\eqref{dgammadx} determines the initial condition $Q(z=0)=Q_0$ with a given $H_0$ due to the relation between $Q$ and $\dot{\gamma}$ . By numerically solving the differential equations with three variables $\gamma,H,Q$ in the MCMC method,  we can constrain the parameter space $\theta=\{\alpha,\beta,\delta,\Omega_m, h, M_B, r_s, \mathcal{X}_0,\mathcal{X}'_0, \lambda,\Omega_k\}$ in the $f(Q)$ theory.

\begin{table*}
\begin{ruledtabular}
\caption{Parameter constraints in three models $\Lambda$CDM, Exp-$f(Q)$, Inv-$f(Q)$ at 1$\sigma$ confidence level (C.L.).}
\label{table1}
\begin{tabular} { cccc ccccc}
Model & Connection & {\boldmath$\Omega_m$} & {\boldmath$h$} &{\boldmath$M_B$} &{\boldmath$\lambda$}& $\mathcal{X}_0$ & $\mathbf{\Omega_k}$  \\
  \specialrule{0em}{5pt}{1pt}
\hline  \specialrule{0em}{10pt}{1pt}
flat $\Lambda$CDM &-&  $0.2898\pm 0.0082$ &$0.696\pm 0.012$ &$-19.367\pm 0.037$ &-&-&- \\
\specialrule{0em}{5pt}{1pt}
non-flat $\Lambda$CDM &-&  $0.489\pm 0.061$ &$0.688\pm 0.012$ &$-19.406\pm 0.040$ &-&-&$-0.39^{+0.11}_{-0.13}$  \\
 \specialrule{0em}{10pt}{1pt}
Exp-$f(Q)$ &$\Gamma_1$& $0.3394\pm 0.0078$ &$0.690\pm 0.012$ &$-19.395\pm 0.036$ &-&-&- \\
\specialrule{0em}{5pt}{1pt}
~&$\Gamma_2$ &$0.259^{+0.016}_{-0.024}$ &$0.686\pm 0.012$ &$-19.405\pm 0.037$ &$0.227^{+0.012}_{-0.017}$ &$-63\pm 21$ &-  \\
\specialrule{0em}{5pt}{1pt}
~&$\Gamma_3$ &$0.492\pm 0.081$ &$0.686\pm 0.012$ &$-19.405\pm 0.041$ &$0.273^{+0.051}_{-0.071}$ &$-428^{+130}_{-100}$ &-  \\
\specialrule{0em}{5pt}{1pt}
~&$\Gamma_4$ &$0.485\pm{+0.066}$ &$0.688\pm 0.013$ &$-19.405\pm 0.041$ &$0.294^{+0.041}_{-0.050}$ &$-274^{+23}_{-19}$ &$-0.34\pm 0.18$ \\
 \specialrule{0em}{10pt}{1pt}
Inv-$f(Q)$ &$\Gamma_1$ &$0.3841\pm 0.0086$ &$0.687\pm 0.012$ &$-19.409\pm 0.038$ &- &-&- \\
\specialrule{0em}{5pt}{1pt}
~&$\Gamma_2$ & $ 0.398^{+0.028}_{-0.015}$ / $< 0.124$ \footnote{From Fig.~\ref{figC2}, the distribution of $\Omega_m$ can be divided into two parts.} &$0.692^{+0.013}_{-0.012}$ &$-19.383^{+0.042}_{-0.034}$ &$-0.303^{+0.032}_{-0.028}$ &$117.9^{+7.1}_{-6.2} $  &-\\
\specialrule{0em}{5pt}{1pt}
~&$\Gamma_3$ & $0.220^{+0.042}_{-0.024}$ &$0.6915^{+0.0087}_{-0.011}$ &$-19.398^{+0.031}_{-0.035}$ &$0.248^{+0.081}_{-0.060}$ &$-98^{+64}_{-20}$  &- \\
\specialrule{0em}{5pt}{1pt}
~&$\Gamma_4$ &$0.254^{+0.022}_{-0.018}$ &$0.694\pm 0.016$ &$-19.384\pm ^{+0.051}_{-0.046}$ &$0.360^{+0.083}_{-0.071}$ &$-167.8\pm7.9$ &$-0.29^{+0.11}_{-0.14}$  \\
 \specialrule{0em}{10pt}{1pt}
\end{tabular}
\end{ruledtabular}
\end{table*}

\begin{figure*}
  \centering
  \includegraphics[width=13cm,height=12cm]{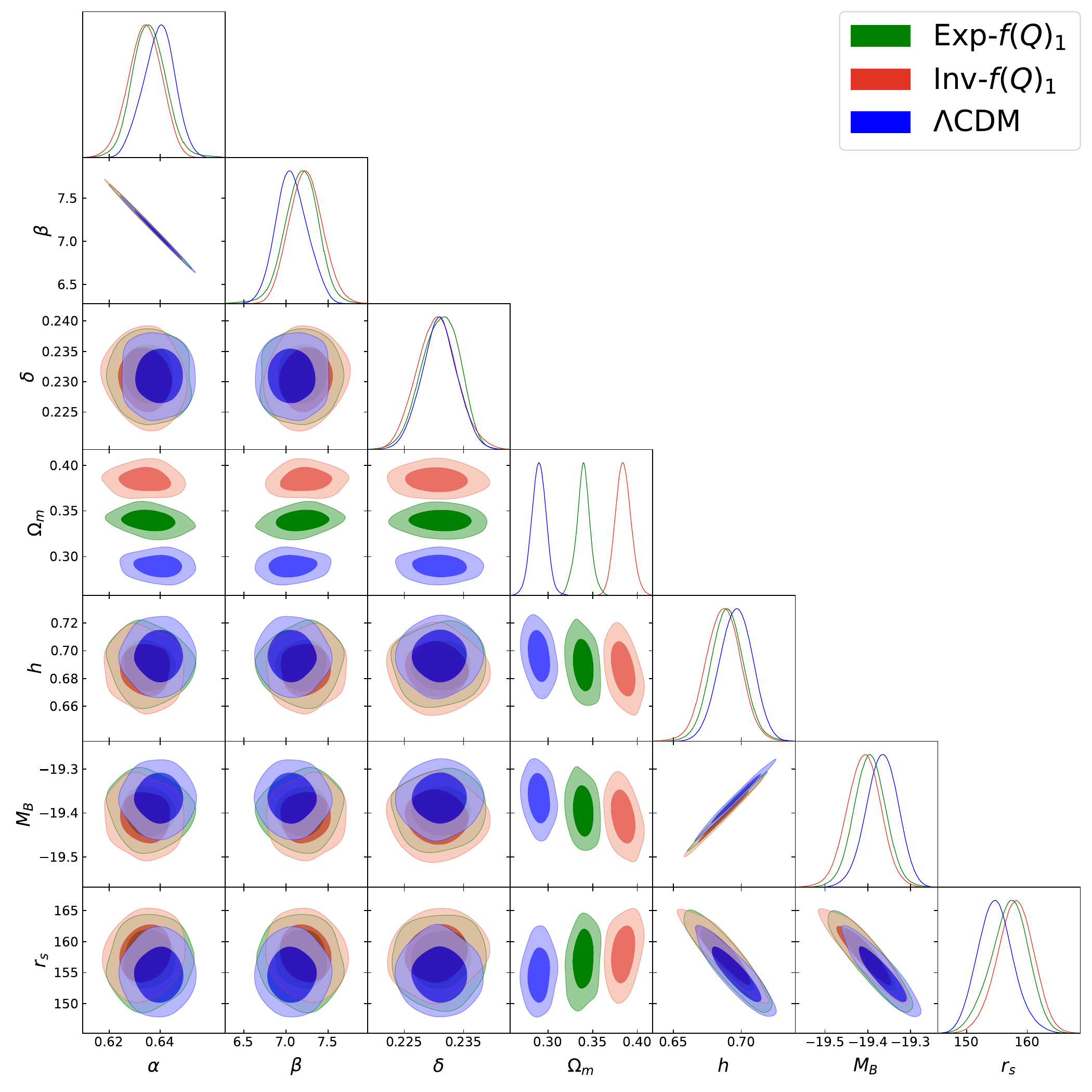}
   \caption{Marginalized constraints at 68\% (darker) and 95\% (lighter) C.L.  on the models with connection $\Gamma_1$ in the flat universe.}
   \label{figC1}
  \includegraphics[width=13cm,height=12cm]{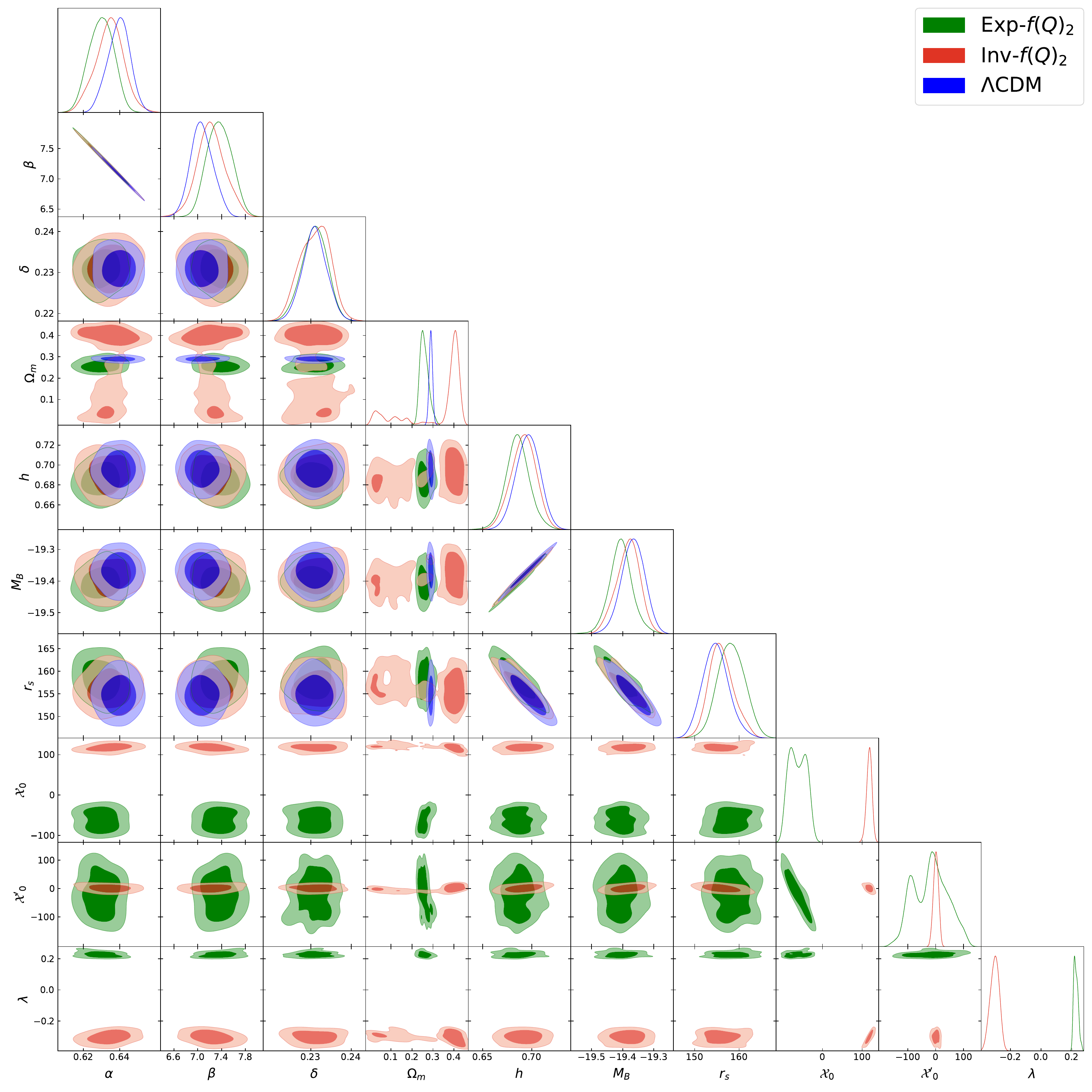}
   \caption{Marginalized constraints at 68\% (darker) and 95\% (lighter) C.L. on the models with connection $\Gamma_2$ in the flat universe.}
   \label{figC2}
  \end{figure*}

  \begin{figure*}
  \centering
  \includegraphics[width=13cm,height=12cm]{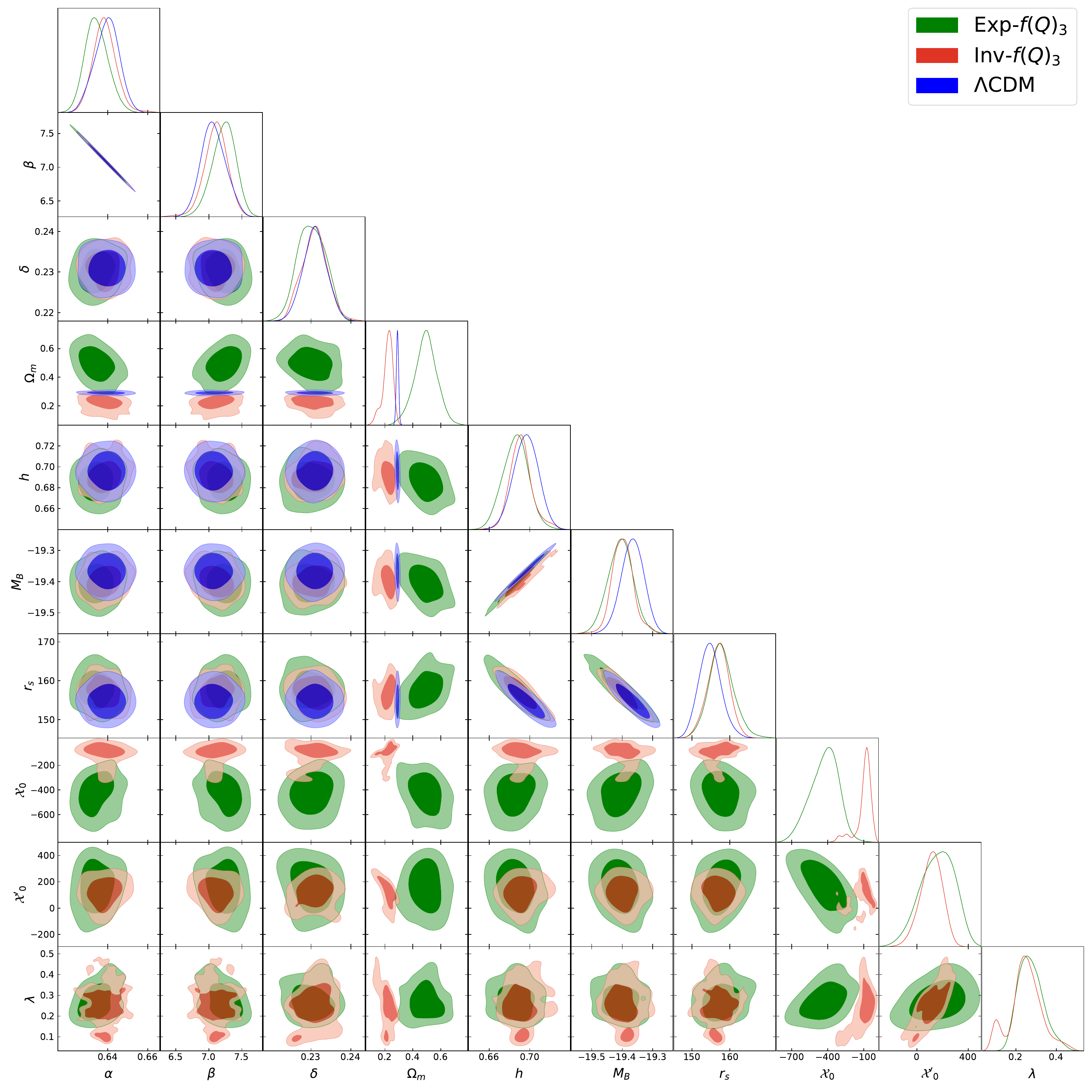}
   \caption{Marginalized constraints at 68\% (darker) and 95\% (lighter) C.L.  on the models with connection $\Gamma_3$ in the flat universe.}
   \label{figC3}
  \includegraphics[width=13cm,height=12cm]{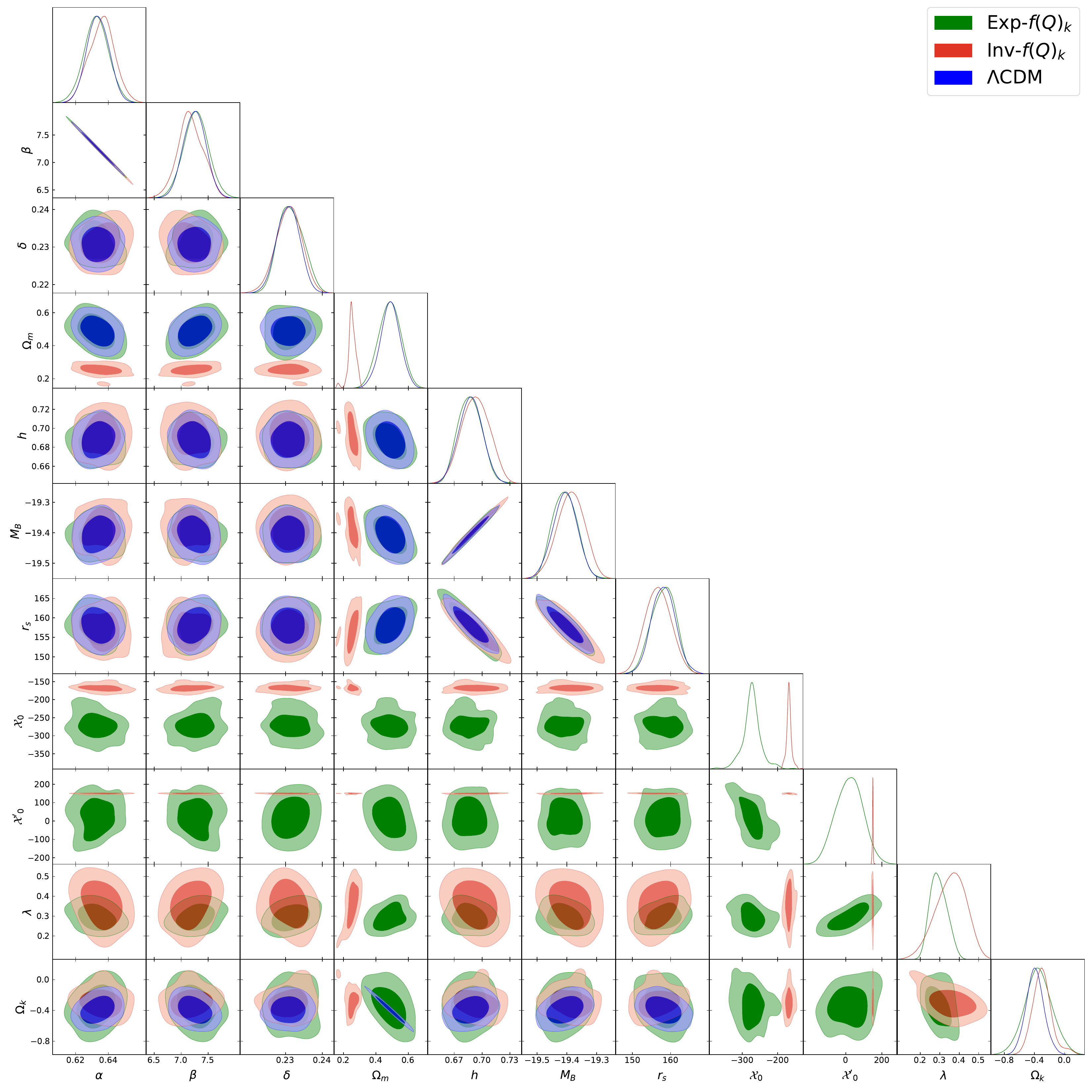}
   \caption{Marginalized constraints at 68\% (darker) and 95\% (lighter) C.L.  on the models with connection $\Gamma_4$ in the curved universe.}
   \label{figC4}
\end{figure*}
\begin{figure*}
  \includegraphics[width=7cm]{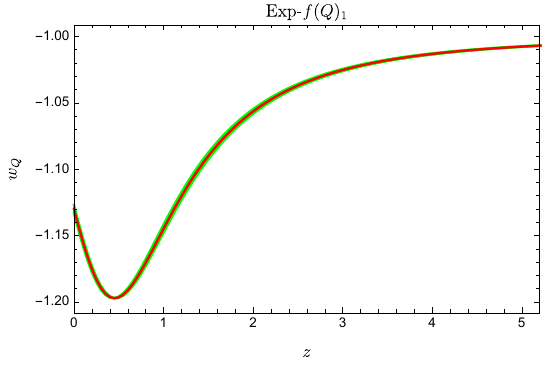}
  \includegraphics[width=7cm]{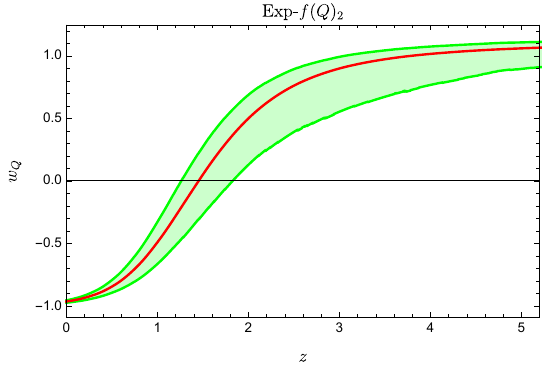}
  \includegraphics[width=7cm]{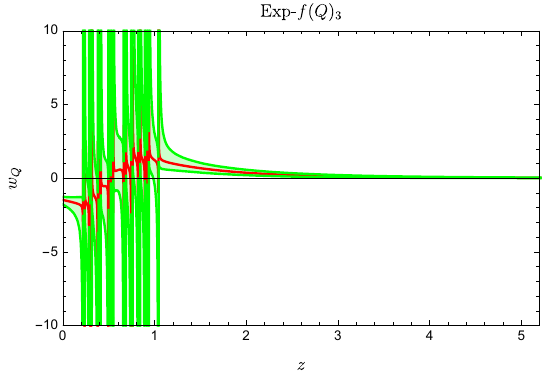}
  \includegraphics[width=7cm]{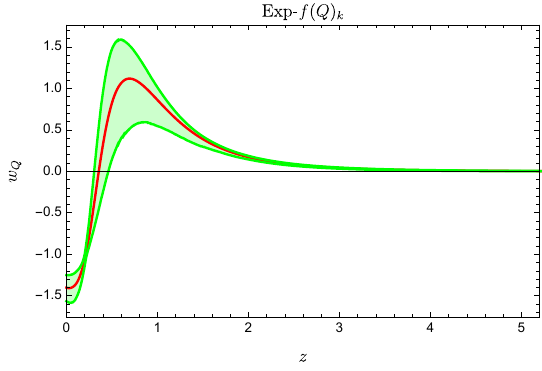}
   \caption{Mean value (red line) and 68\% C.L. regions (green band) for the evolution of the EoS $w_Q$  vs redshift $z$ in Exp-$f(Q)$ models. }
   \label{figw1}
\end{figure*}

\begin{figure*}
  \includegraphics[width=7cm]{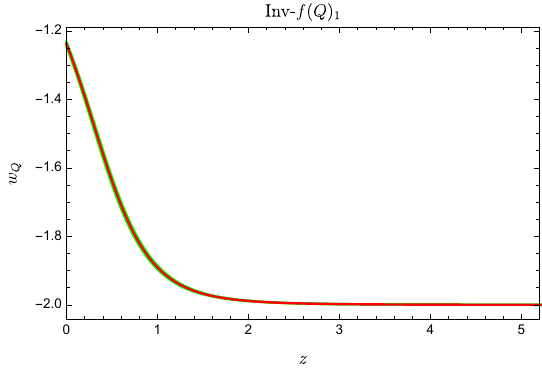}
  \includegraphics[width=7cm]{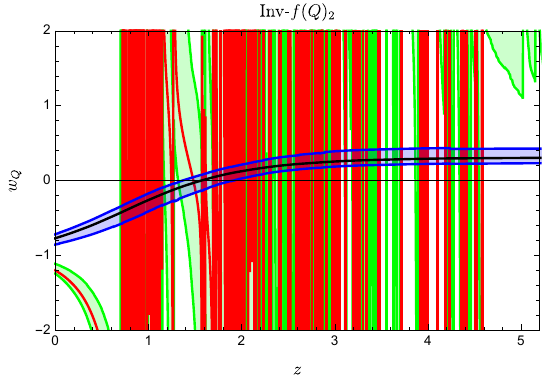}
  \includegraphics[width=7cm]{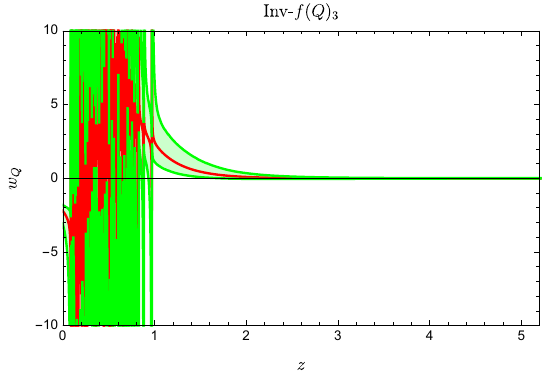}
  \includegraphics[width=7cm]{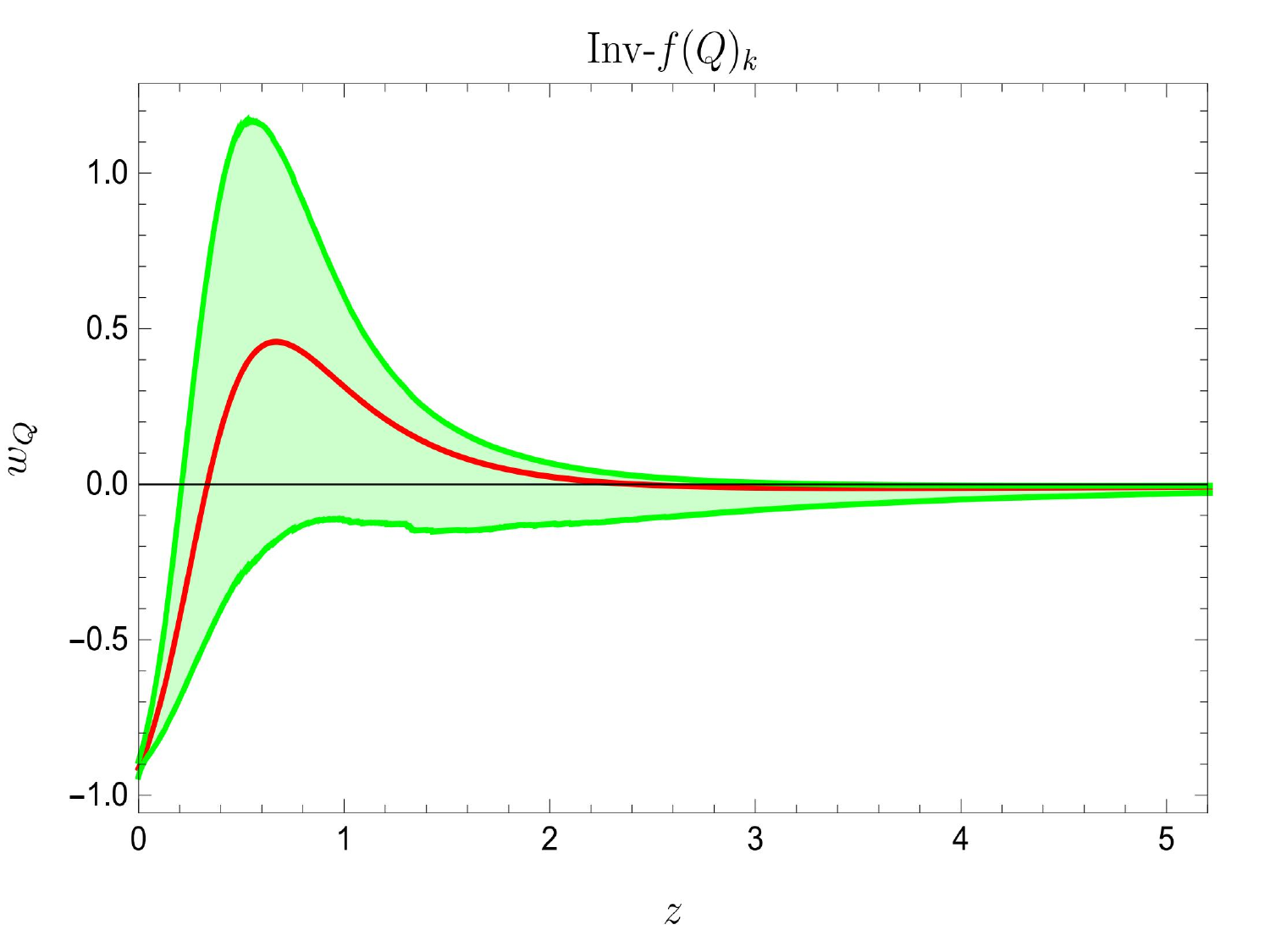}
   \caption{Mean value (red/black line) and 68\% C.L. regions (green/blue band) for the evolution of the EoS $w_Q$  vs redshift $z$ in Inv-$f(Q)$ models. For the Inv-$f(Q)_2$ model, the distribution of $\Omega_m$ can be divide in two parts as shown in Fig. \ref{figC2}. The blue band of $w_Q$ corresponds to the case of $\Omega_m<0.25$ and the green band corresponds to the case of $\Omega_m\geq0.25$.}
   \label{figw2}
\end{figure*}

In order to find the posterior distribution of the parameters in the MCMC method, a Python module \texttt{emcee} \cite{Foreman-Mackey:2012any} is used to produce MCMC chains, and the \texttt{GetDist} \cite{Lewis:2019xzd} package is used for statistic analysis and the plotting of posterior probability distributions of the parameters. For each affine connection in the $f(Q)$ theory, I denote $f(Q)_i$ with the subscript $i$ representing the models with different connection $\Gamma_i$ and use $f(Q)_k$ to represent the case of spatial curvature. Here the total $\chi^{2}$ used in the MCMC algorithm is
\begin{equation}\label{tchi}
\chi_{\mathrm{Total}}^{2}=\chi_{\mathrm{SNe}}^{2}+\chi_{\mathrm{CC}}^{2}+\chi_{\mathrm{BAO}}^{2}+\chi_{\mathrm{QSO}}^{2}\,.
\end{equation}
The median values of model parameters and 1$\sigma$ confidence intervals on them are presented in Table \ref{table1}. The contour plots of the posterior probability distribution of the parameters are shown in Figs. \ref{figC1}-\ref{figC4}. The evolutions of $w_Q$ vs redshift $z$ are displayed in Fig. \ref{figw1} and \ref{figw2} corresponding to the models Exp-$f(Q)$ and Inv-$f(Q)$, respectively.

Summarizing the aforementioned table and plots, some interesting information is highlighted as follows:

(i) For the $f(Q)_1$ models shown in Fig. \ref{figC1}, the values of $\Omega_m$ in the $f(Q)$ theories have significant deviations from the concordance $\mathrm{\Lambda}$CDM model, which is in agreement with the result in Ref. \cite{Anagnostopoulos:2021ydo}. $f(Q)$ cosmology, to some extent, challenges the result from the $\mathrm{\Lambda}$CDM model as later I will compare the Information Criterion (IC) of various models. Moreover, these deviations also happen between Exp-$f(Q)$ models and Inv-$f(Q)$ models. From Figs. \ref{figC2}-\ref{figC4}, such deviations on $\Omega_m$ also slightly occur in $f(Q)_2$, $f(Q)_3$ , $f(Q)_k$  models except for the case of the Exp-$f(Q)_k$ and the $\Lambda$CDM.

(ii) Unlike the models with other connections, the parameter $\lambda$ is not free but dependent on $\Omega_m$ and $\Omega_r$ in the $f(Q)_1$ models. Using the MCMC chains of $\Omega_m$ and $\Omega_r$, the parameter $\lambda$ can be determined as $\lambda = 0.3669\pm 0.0034$ in the Exp-$f(Q)_1$ model and
$\lambda = 0.2053\pm 0.0029$ in the Inv-$f(Q)_1$ model. This implies no extra parameters introduced in the $f(Q)_1$ models. As a side note, the positive $\lambda>0$ is preferred as the condition for an attractor in Exp-$f(Q)_1$ cosmology though the negative $\lambda$ is also the solution \cite{Ferreira:2023kgv}.

(iii) All the Exp-$f(Q)$ and Inv-$f(Q)$ models favor a nonzero $\mathcal{X}_0$ as its peak of posterior distribution departs from $\mathcal{X}_0=0$. That means the inertial effects are not negligible in the cosmic history.

(iv)  A negative value of curvature density parameter is favored in the $\mathrm{\Lambda}$CDM model: $\Omega_k=-0.39^{+0.11}_{-0.13}$ , which indicates that the universe is closed. As a matter of fact, the evidences for the closed universe in the $\mathrm{\Lambda}$CDM model were also verified by the higher redshift data including Planck 2018 observations \cite{Planck:2018vyg,DiValentino:2019qzk,DiValentino:2020hov,Handley:2019tkm,Yang:2022kho} and QSO observations \cite{Wei:2019uss,Khadka:2020tlm,Khadka:2021xcc,Bargiacchi:2021hdp,Dinda:2023svr}. Especially, both model dependent and independent methods favor a closed Universe when using QSO data \cite{Dinda:2023svr}.
This situation of a closed universe also happens in Exp-$f(Q)_k$ and Inv-$f(Q)_k$ models as shown in Fig. \ref{figC4}.
  By the way, the non-flat $\mathrm{\Lambda}$CDM model predicts a higher matter density $\Omega_m\sim 0.5$ in striking contrast with local measurements of galaxy clustering. This result is in agreement with the constraints from the observations of cosmic microwave background (CMB) by Planck 2018 \cite{DiValentino:2019qzk}. In fact there is an increasing $\Omega_m$ trend with the redshift in the flat universe as discussed in \cite{Colgain:2022nlb,Malekjani:2023dky}. The introduction of extra parameter $\Omega_k$ probably affects the emergence of this trend, which needs further investigations. However in the Inv-$f(Q)_k$ model, a lower value $\Omega_m=0.277\pm 0.045$ is given which is different from the result in the $\mathrm{\Lambda}$CDM model and the Exp-$f(Q)_k$ model.

(v) The evolutions of dark energy EoS $w_Q$ are shown in Fig. \ref{figw1} and Fig. \ref{figw2}, where the mean value of $w_Q$ (red line) and its 1$\sigma$ C.L. region (green band)  are displayed. It can be easily found the phantom crossing behavior happens in the Inv-$f(Q)_2$, Exp/Inv-$f(Q)_3$ and Exp-$f(Q)_k$ models, which is absent in $f(Q)_1$ models. However the Exp-$f(Q)_2$ and Inv-$f(Q)_k$ models does not have such a significant phantom crossing behavior. For the Inv-$f(Q)_2$ in Fig. \ref{figw2}, the phantom crossing behavior disappears in the blue band of $w_Q$ corresponding to the case of $\Omega_m<0.25$. Using Eqs. \eqref{modF21} \eqref{wQ}, quite a large amplitude of $w_Q$ in the figures indicates the appearance of the negative dark energy density $\rho_Q<0$ in the Exp-$f(Q)_3$, Inv-$f(Q)_2$ and Inv-$f(Q)_3$ models.

\subsection{ Model comparison}
\label{dataC}

\begin{table*}
\caption{Summary of the $\chi^2_{\mathrm{min}}$ values and various information criteria for the cosmological models (flat case $k=0$).}
\label{table2}
\begin{tabular} { ccccc ccc}

\hline \hline  \specialrule{0em}{5pt}{1pt}
Model & Connection & $\Delta \chi^2_{\mathrm{min}}$& $\Delta$AIC& $\Delta\mathrm{ AIC_c}$ &$\Delta$BIC &$\Delta$DIC &$\mathcal{B}_{ij}$  \\
  \specialrule{0em}{5pt}{1pt}
\hline  \specialrule{0em}{10pt}{1pt}
flat $\Lambda$CDM &-&  0 &0 &0&0&0&1 \\
\specialrule{0em}{10pt}{1pt}
Exp-$f(Q)$ &$\Gamma_1$&  -4.193 &-4.193&-4.193 &-4.193 &-3.821&7.8   \\
\specialrule{0em}{5pt}{1pt}
 ~&$\Gamma_2$&   -4.508& 1.492& 1.532& 19.185& -19.607 &4.0  \\
\specialrule{0em}{5pt}{1pt}
~&$\Gamma_3$&-4.693& 3.307& 3.364& 26.898& -3.644 &6.0   \\
\specialrule{0em}{5pt}{1pt}
Inv-$f(Q)$ &$\Gamma_1$&-4.789 &-4.789& -4.789 &-4.789 &-4.851&11.4 \\
\specialrule{0em}{5pt}{1pt}
 ~&$\Gamma_2$& -2.567 &3.433& 3.474& 21.126& -9.357&2.2 \\
\specialrule{0em}{5pt}{1pt}
~&$\Gamma_3$&-4.799& 1.201& 1.241& 18.894& -10.728&5.2\\
\specialrule{0em}{5pt}{1pt}
\hline \hline
\end{tabular}
\end{table*}

\begin{table*}
\caption{Summary of the $\chi^2_{\mathrm{min}}$ values and various information criteria for the cosmological models (non-flat case).}
\label{table3}
\begin{tabular} { cccccc cc}

\hline \hline  \specialrule{0em}{5pt}{1pt}
Model & Connection & $\Delta \chi^2_{\mathrm{min}}$& $\Delta$AIC& $\Delta\mathrm{ AIC_c}$  &$\Delta$BIC &$\Delta$DIC &$\mathcal{B}_{ij}$  \\
  \specialrule{0em}{5pt}{1pt}
\hline  \specialrule{0em}{10pt}{1pt}
non-flat $\Lambda$CDM &-&  0 &0 &0&0&0&1 \\
\specialrule{0em}{10pt}{1pt}
Exp-$f(Q)$ &$\Gamma_4$&0.206& 6.206& 6.251& 23.899& 1.954 &0.7\\
\specialrule{0em}{10pt}{1pt}

Inv-$f(Q)$ &$\Gamma_4$&0.253 &6.253& 6.298& 23.946 &0.536& 0.5 \\

\specialrule{0em}{5pt}{1pt}
\hline \hline
\end{tabular}
\end{table*}

Some information criteria are widely used in astrophysics and cosmology to compare various models for the evidence, including the Akaike Information Criterion (AIC) \cite{1100705}, the corrected Akaike Information Criterion ($\mathrm{AIC_c}$) \cite{doi:10.1080/03610927808827599}, the Bayesian Information
Criterion (BIC) \cite{10.1214/aos/1176344136} and the Deviance Information Criterion \cite{10.1111/1467-9868.00353}
(DIC), defined as follows \cite{Liddle:2007fy}
\begin{align}\label{AIC1}
&\mathrm{AIC}=\chi_{\mathrm{min}}^{2}+2k \,,\\
\label{AIC2}
&\mathrm{AIC_c}=\mathrm{AIC}+\frac{2k(k+1)}{N-k-1}\,,\\
\label{BIC}
&\mathrm{BIC}=\chi_{\mathrm{min}}^{2}+k\ln N\,,\\
\label{DIC}
&\mathrm{DIC}=D(\overline{\theta})+2p_{D}\,,
\end{align}
where $\chi_{\mathrm{min}}^{2}$ is the minimum value of chi-square, $k$ is the number of free parameters and $N$ is the total number of data points in the data combinations. $D$ is the deviance of the likelihood for the parameter space $\theta$, i.e. $D(\theta)=\chi^{2}(\theta)+C$ with $C$ as a constant. The effective number of parameters in the model is $p_{D}=\overline{\chi^{2}(\theta)}-\chi^{2}(\overline{\theta})$. By using this effective number of parameters, the DIC overcomes the problem of the AIC and BIC that they do not discount parameters that are unconstrained by the data. However, AIC and BIC are reasonable to evaluate the evidence level of models if the parameters in the models are constrained well respecting the Gaussianity of the posterior distribution. Moreover, $\mathrm{AIC_c}$ is used for small sample sizes while $\mathrm{AIC}$ is its limit value as $N\gg k$.  I choose the $\mathrm{\Lambda CDM}$ model as the reference model to define IC differences
$\Delta\mathrm{IC}=\mathrm{IC}(\mathrm{model})-\mathrm{IC}(\mathrm{\Lambda CDM}),\mathrm{IC=AIC,AIC_c,BIC,DIC}$. Given the value $\Delta\mathrm{IC}$, the evidence level of the model can be evaluated. As a general rule of thumb,  one usually considers  $\Delta\mathrm{IC}<2$ to indicate substantial support (evidence), $4<\Delta\mathrm{IC}<7$  much less support,
and $\Delta\mathrm{IC}>10$  essentially no support \cite{delaCruz-Dombriz:2016bqh,doi:10.1177/0049124104268644}.

As a direct way to compare different models, the Bayesian evidence can tell which model is more favored by observations. It comes from a full implementation of Bayesian inference at the model level. The Bayes factor of model $\mathcal{M}_{i}$ with respect to model $\mathcal{M}_{j}$ is given by
\begin{equation}\label{bf1}
\mathcal{B}_{ij}=\frac{P(D|\mathcal{M}_{i})}{P(D|\mathcal{M}_{j})}\,,
\end{equation}
here the Bayesian evidence is
\begin{equation}\label{bf2}
P(D|\mathcal{M}_{i})=\intop d\theta_{i}p(\theta_{i}|\mathcal{M}_{i})\mathcal{L}(D|\theta_{i},\mathcal{M}_{i})\,,
\end{equation}
where $p(\theta_{i}|\mathcal{M}_{i})$ is the prior probability for the parameters $\theta_{i}$, and $\mathcal{L}(D|\theta_{i},\mathcal{M}_{i})$ is the likelihood of the data $D$ given the model parameters $\theta_{i}$. A Bayes factor $\mathcal{B}_{ij}>1$ indicates that model $\mathcal{M}_{i}$ is more strongly supported by data than model $\mathcal{M}_{j}$. When $1<\mathcal{B}_{ij}<3$ there is evidence against $\mathcal{M}_{j}$ when compared with $\mathcal{M}_{i}$, but it is only worth a bare mention. When $3<\mathcal{B}_{ij}<20$ the evidence against $\mathcal{M}_{j}$ is definite but not strong. For $20<\mathcal{B}_{ij}<150$ the evidence is strong and for $\mathcal{B}_{ij}>150$ it is very strong \cite{Drell:1999dx,John:2002gg,Nesseris:2012cq}. In this paper,  the $\Lambda$CDM  model is fixed as the fiducial model $\mathcal{M}_{j}$.

The model comparison results are listed in Table \ref{table2} for the flat universe and Table \ref{table3} for the non-flat universe. From Table \ref{table2}, the $f(Q)$ models are strongly supported on the whole except for the AIC/AICc and BIC of the $f(Q)_2$ and $f(Q)_3$ models in contradiction to other model selection methods.  The emergence of the conflict between the different selection methods is not strange \cite{Rezaei:2021qpq}. From Table \ref{table3}, all the Information Criteria and the Bayes factor indicate the less support for the $f(Q)_k$ models. Judging from the characteristics of these model selection methods, the DIC and the Bayesian evidence are more reliable to compare models.  Based on the results from the DIC and the Bayesian evidence, it is safe to claim that the $f(Q)$ models with all different connections in the flat universe are more favored compared to the $\Lambda$CDM  model, while the $\Lambda$CDM model in the non-flat case is more favored than the $f(Q)$ models in each model comparison method. Furthermore, the non-trivial connections in $f(Q)_2$ and $f(Q)_3$ have less support evidence than the connection $\Gamma_1$ based on the Bayesian evidence. That means the coincident gauge in $f(Q)$ theory is the best gauge for matching the cosmological observations. It indicates that the absence of the inertial effects in the flat universe is more favored.

\section{Conclusions and discussion}
\label{Conclusions}

In this paper, the cases of flat and non-flat universes are investigated with the observational date sets SNe+CC+BAO+QSO in the covariant $f(Q)$ theory. Two specific $f(Q)$ models with different affine connections are proposed to be examined by cosmological observations. The $f(Q)$ theory offers a component of effective dark energy to alleviate the cosmological constant problem.   The non-trivial connections have the support evidence compared to the $\Lambda$CDM model in the flat universe based on the model selection methods of the DIC and the Bayes factor. Moreover, the coincident gauge or the connection $\Gamma_1$ is the best gauge choice consistent with observations, which supports the zero inertial effect in the flat universe. Besides this,  the $\Lambda$CDM model has higher support evidence level than the $f(Q)$ models in the non-flat universe. In the non-flat case, the negative values of $\Omega_k$ are given in the $\Lambda$CDM model and the $f(Q)$ models.  It is consistent with the spatial curvature constraints as the hints of a closed universe from the QSO observational data via the model dependent and independent analyses \cite{Dinda:2023svr}. Finally the non-trivial connections in the Inv-$f(Q)_2$, Exp/Inv-$f(Q)_3$ and Exp-$f(Q)_k$ models can induce a significant phantom crossing  behavior which is absent in the other connections. Such a situation can be regarded as the gauge induced phantom crossing (GIPC) behavior. Here the support evidence for the $f(Q)$ dynamical dark energy is quite strong, in agreement with the evidence from the other cosmological observations \cite{Zhao:2017cud,Capozziello:2018jya}. As a side note, the support evidence level of Inv-$f(Q)$ compared to the $\Lambda$CDM model in this paper is contradictory to \cite{Ayuso:2020dcu}, which may be attributed to the usage of the distinct data sets.

Here two $f(Q)$ models are given, we can also extend the study in other $f(Q)$ models, such as the power law form $f(Q)\varpropto Q^n$ or other more complicated forms as shown in \cite{Arora:2022mlo,Anagnostopoulos:2022gej}. It is very intriguing to investigate whether the non-trivial connection can induce the effective dark energy or not while the connection $\Gamma_1$ can not in some specific models. If the non-trivial connection contributes an effective dark energy, such an effective dark energy coming from the inertial effects will be absent in the inertial frame.

Many more possibilities of $f(Q)$ models have not yet been investigated with observational data when $\gamma$ is an arbitrary function of time $t$. It is also interesting to study the evolution of the function $\gamma$ in an unknown $f(Q)$ model.  For example, the case of $\gamma\varpropto a(t)$ has been discussed in \cite{Subramaniam:2023old,Shabani:2023xfn} and even the theory with a general time-varying $\gamma(t)$ is studied by phase-space analysis \cite{Shabani:2023nvm}. The investigation of cosmological constraints on the function $\gamma$ without the explicit expression of $f(Q)$  will be presented in coming work.
\begin{acknowledgements}
I wish to thank the anonymous referee for important and helpful comments. J.S. was partially supported by the Fundamental Research Funds for the Central Universities (Innovation Funded Projects) under Grants No.~2020CXZZ105.
\end{acknowledgements}

\bibliographystyle{apsrev4-1}
\bibliography{fQ202306}

\end{document}